# Navigating Incommensurability Between Ethnomethodology, Conversation Analysis, and Artificial Intelligence

Stuart Reeves
School of Computer Science, University of Nottingham, UK

Like many research communities, ethnomethodologists and conversation analysts have begun to get caught up—yet again—in the pervasive spectacle of surging interests in Artificial Intelligence (AI). Inspired by discussions amongst a growing network of researchers in ethnomethodology (EM) and conversation analysis (CA) traditions who nurse such interests, I started thinking about what things EM and the more EM end of conversation analysis might be doing about, for, or even with, fields of AI research. So, this piece is about the disciplinary and conceptual questions that might be encountered, and—in my view—may need addressing for engagements with AI research and its affiliates. Although I'm mostly concerned with things to be aware of as well as outright dangers, later on we can think about some opportunities. And throughout I will keep using 'we' to talk about EM&CA researchers; but this really is for convenience only—I don't wish to ventriloquise for our complex research communities. All of the following should be read as emanating from my particular research history, standpoint etc., and treated (hopefully) as an invitation for further discussion amongst EM and CA researchers turning to technology and AI specifically.

**We Have Been Here Before**
Why do I feel the need for some caution? Well, *we have been here before*, most recently in the 1990s with works by Button et al. (1995) and Luff et al. (1990). We need to assess what happened then, and how this relates to now. For instance, there were greater technical limitations in the 80s and 90s which are far less present now—was this a reason for ECMA research on AI going mostly dark since then? Or owing to the vagaries of research funding?

A second initial framing question is that EMCA researchers' options for engagement with 'AI' (whatever that is—we'll discuss that problem shortly) are broader perhaps than they used to be. Engaging with AI does not *necessarily* mean 'direct' coupling with that particular discipline at all. Renewed interest in AI has surfaced very strongly—this time at least—in research communities with stated interests in 'humans' (again, whatever that means). The most obvious is human-computer interaction (HCI) research, with which AI shares a long dialectical history (Grudin 2017), although I imagine more HCI researchers are aware of this than AI ones given their relative prominence on the intellectual landscape. There are also more specialist fields which have a strong existing affiliation to AI, like human-robot interaction (HRI) or intelligent user interfaces (IUI), which benefit from the last ~10 years of AI advances, and which also may act as potential recipients of EMCA work on AI. What does it mean for EMCA oriented work to address AI within the confines of such fields?

These are both very initial matters to consider. But beyond this I think there are further questions which run deeper into what the (likely different) programme(s) and emphases of EM and CA in their engagements with AI might turn into. In other words, what is EMCA research *doing* with AI and what is it meant to be *for*? Let's start with some potential dangers.

**Some Dangers We Might Heed**
Some immediate matters for EMCA—and, perhaps for anyone trying to take AI unironically—encapsulated by both the name 'artificial intelligence' and the very idea: fundamentally how do you avoid a naive approach? To begin, 'artificial intelligence' is one of the few disciplinary designations I can think of which is *aspirational*—not necessarily a very good sign for solidity. The name is an original sin of the field, dreamt up for the [Dartmouth workshop in 1956 by John McCarthy](), or so the story goes. The effect of this name has been both egregiously misleading and at the same time invigorating for the communities cleaving to it. AI researchers—at least the disciplinarily competent—*tend to know what game they are playing* and thus—I believe—can readily 'point' in concert with most fellow travellers to this thing or that thing as 'AI' (this is not to suggest an absence of debate, of course). The language nevertheless generates immense trouble: it can be advantageous for researchers to disown examples of AI endeavours which are not so impressive in the end, no longer call them 'AI', and lock them out of the disciplinary association. Instead, strategically minded AI researchers will take ownership over any recent successes that hype the discipline: hence, the most salient thing we can say about DeepMind's Go-playing system [AlphaGo]() (Mair et al. 2021), for instance, is that it is a public demonstration of 'AI' that—just like IBM's Deep Blue example—simultaneously does all manner of work for organisations, funding, branding, media coverage, etc.

At the same time as such public spectacles of AI, widespread adoption and normalisation of technologies which also have a legitimate lineage rooted in AI research are often subsequently no longer identified as AI—they simply are absorbed into the technological plenum. This is just another side effect of AI's aspirational framing: when you are no longer aspirational but here and now, you are no longer part of the conversation. Technologies like spam filtering, auto-correct or auto-completion are all 'AI' (credit to Barry Brown for this observation), but the discourse rarely concerns itself with them; AI is instead AlphaGo, self-driving cars, or domestic robots. This makes me wonder *what* EMCA research can even grasp on to *as 'AI'*, without looking foolish or gullible. By what definition, vernacular or otherwise could we or should we even identify technologies as AI, if at all? I constantly wonder if there is anything fundamentally different about AI technologies phenomenologically when categorical boundaries are so fluid. Maybe we should look at those boundaries—bracketing *what counts as AI*—as social accomplishments. In any case, this first issue for EMCA's relations to AI is that living in someone else's imagination can be dangerous place (especially if it is a fever dream of, say, self-driving cars).

For EMCA research to begin to handle these challenges, I feel it might also need deal with the mystification of what I'll call 'agentification' (Reeves and Porcheron 2022). By this I mean the reification of the technological object as a marvel, thoroughly detached from the circumstances of its production (both in a manufacturing sense and sociological sense). I'm not sure this concept is novel, but my neologism maybe has some utility. Agentification is enacted, worked upon, and sustained by researchers forming communities of care for that technological object, practices which themselves further agentify AI because they are then erased. This is a fundamental dynamic that I'm sure is true of most technologies, but perhaps becomes clearest for AI. The same goes for constellations of active users, people implicated in the AI system, and so on. The effect of this has been significant confusion about what is, what is not AI, what (social, technical) constituent 'parts' are relevant, etc. (see Mair et al. 2021). For ethnomethodologists and conversation analysts, it might be that attending to such categorical work done by members yields enlightening (and likely somewhat deflating) studies 'of AI'. (This work might also need to take stock of the notion that as a confection, 'AI' disappears the more you reify it.)

**Understanding the Cauldron**
This last point brings me to discuss another way to talk of AI which could help attune EMCA research approaching AI spaces: this is in thinking about AI as a cultural phenomenon; i.e., as a complex amorphous community of researchers and practitioners crossing industry and academia. Perhaps an obvious point, but I think it has implications for EMCA's potential engagements with AI. There is not so much a porous boundary between these as a fluid set of overlapping identities and motivations. Fields of AI consist of a variety of specialisms that mostly stem intellectually and culturally from computer science including robotics, and engineering. But there are also strong intellectual and cultural affinities with cognitive science, crossovers with neuroscience, and traditions within philosophy, that are at play. We might need to be cognisant of these: they range from explicitly accounted-for motivating agenda, through to something more like a 'background radiation'; all, sets of normative practices for researchers within AI, as touched upon by Forsythe (2002).

There are also 'clients' which adopt AI techniques that I alluded to earlier, like HCI, and maybe in some senses HRI and IUI communities, which operationalise AI technologies in some way or another but also have elements that more directly participate in AI communities. Yes, there is a bit of talkback but it's generally clear where the core agenda (e.g., building novel AI techniques) tend to be driven from when you are mostly a 'client'.

Finally there are communities of 'critics' (using this word advisedly because many of these may well be 'within' core AI communities). These tend to examine the claims of AI and either focus on questions of tractability, feasibility, or perhaps concentrate on understanding AI's sociotechnical implications. This simplistic version I've given is the what we could call 'communities of AI-affiliate research'.

Obviously this is an incomplete description. But the point still stands: if EMCA research(ers) want to engage, where to engage, and what's the strategy?

Understanding these communities means also understanding the discourse of artificial intelligence. Note that *in no way* am a historian or an expert in AI's history; rather, I just want to highlight the need for appreciating this discourse's history. It extends most obviously back to the 50s and 60s (although has clear antecedents earlier than this which I won't go into here). I tend to think there are a few paradigms or modes of this which we can discern, from an outsider's perspective (and my limited one at that).

Let's consider a very initial, incomplete list that crassly characterises views (sorry).

The most obvious is the singularity people—these are accelerationists for AI, who believe that computational versions of intelligence are a done deal, we just need the technology to advance enough. This view, I think, relies substantively on self-delusion in spite of a long list of fairly obvious fundamental problems (Smith 2018).

There are centrists, maintaining that the concept of 'intelligence' is still relevant to AI, but that it's nevertheless a tough nut to crack. For them, maybe we won't get there—intelligence is possibly ineffable—but the general direction of travel remains promising.

There are 'inner' critics, a subgroup perhaps of centrists, such as Marcus (2019) who argues that we need different approaches, paradigm shifts, to develop true (useful) AI (Mitchell 2019), whilst maintaining a strong sense of skepticism over machine learning-related advances. The relevance of concepts of 'intelligence' are thus still a feature of AI centrism. Some of these people (not Marcus or Mitchell I imagine) are what I'd call data-world isomophism-ists—that is, a belief that 'data'—AI fuel—ultimately can sufficiently represent the world, people, or whatever it might be, in enough detail so as to enable the aspirations of AI. It might be that a reorienting of AI's direction towards truly understanding something like 'common sense' is part of such a route forwards, we just need the data in some way.

Besides these singulatarians and centrists there are the utilitarians who take a leave of absence for anything beyond an immediate entirely technical conceptualisation of AI problems, presumably tracing little intellectual connection to Turing, 1950s AI, and the emergence of stubbornly persistent preoccupations around artificial general intelligence (AGI), Turing Tests, and so on. The perspective is: do not question foundational concepts, just keep picking the fruits. In a sense, utilitarian AI research can on the one hand laud metric improvements predicated on things like 'human performance' equivalents whilst simultaneously remaining conveniently distant from the debates over what is meant by 'intelligence' in AI. (There is no 'neutral' position, of course, but that is another matter.)

Further along we have dissenters, who may argue that something is fundamentally missing from AI as a programme—it is deficient in its base conceptions. Whilst this position *could* be a centrist one, the argument tends not to be 'we just need better techniques in AI' or that 'AI is broken but can be fixed', but rather that the field itself is foundationally incomplete or at a dead end, because it does not appreciate some set of key issues. Once again, this dissent works partly because of the aspirational quality of 'artificial intelligence' which really can be taken as a purely technical term through to something philosophical, mystical, or even perhaps as a manifesto. For instance, Dreyfus—at the time reacting to the inflated claims of AI research in the 1960s and 70s—pointed out that various fundamentals were wholesale missing from (symbolic) AI, like notions of embodiment or an understanding of tacit knowledge (Dreyfus 1992).

Related to dissenters but probably distinct from it are EM grounded critiques of Button et al. (1995). The difference here, I think, which EM brings is that they amount to a critique of the presumptions of AI research achieving 'intelligence' as 'not even wrong', i.e., a category mistake in Ryle's terms. For instance natural language understanding (NLU) is little about language or understanding—it's certainly about something called those things but they are not commensurate with how such terms are normally used. An EM critique (in my mind) instead points to understanding AI technologies like NLP as part of a members' achievement—of design, programming, staging, demonstration, etc. (Mair et al. 2021). This critique covers both approaches: symbolic (Good Old Fashioned AI—GOFAI) and subsymbolic (e.g., machine learning, but more strongly neural networks). Relatedly we could point to Shanker's work on Wittgensteinian perspectives on AI which makes similar distinctions (Shanker 1998).

If there *is* an EMCA that intersects with AI somehow, how will it navigate this cauldron of discourses? And how might EM and CA researchers working with AI-driven technologies position themselves given prior critiques by Button et al. and others? I don't have any good answers but I do have those questions at least.

I want to close this section with a few broad comments. Firstly, Button et al. have distinguished themselves—extensively and in a detailed way—from prior critiques of AI mythologies or imaginations, whether that is Dreyfus or Searle, and advanced some clear moves against the stack of core conceptual confusions encased within AI programmes of the past, many of which are little different intellectually to the present. Strategically for EM and CA, it might be that new technological circumstances reveal themselves as opportunities for restating these fundamental criticisms and conceptual concerns for new audiences. Or they may reveal novel phenomena not really available to Button et al. at the time. I remain a bit hesitant because (in my view) much of what we had to say about 'conversational AI' systems (e.g., Porcheron et al. 2018) felt indebted strongly to rerunning both Suchman's and Button et al.'s arguments.

Secondly, there are real potential harms driven by AI technologies, including the (mis)application of alleged 'recognition' technologies to social circumstances, or other data-driven AI techniques like those based in machine learning. Fields like Science and Technology Studies, policy and law research, coupled with the emergence of organisations addressing these topics (e.g., AI Now Institute), and the seemingly endless flow of AI ethics oriented investigations that address matters like regulation or the general 'social implications of AI'—these seem to have a lot of those troubles with AI on the books already. This has led to an increasing sensitisation of many research communities in harms related to AI and even an occasionally-critical or skeptical media following suit. This push has been relatively successful in at least starting to raise awareness (if not some action) of / on e.g., machine-driven bias, or attempts to fight against the absolution of responsibility for AI systems via the 'agentification' trick (e.g., many confused discussions around self-driving cars and 'responsibility'). These critical perspectives are, very gradually, succeeding in foregrounding the social / cultural / societal / political implications of AI both within research and practitioner communities. But the question here is then about where EMCA research fits in amongst the various fighting fronts—and what is its potential role? I feel that the best I can come up with is to remind myself of EMCA's relentless concern with concrete actualities of social order—order that is enacted, produced around and taking into account AI systems in general. Repeating a very common refrain from EMCA research, this view of sociality as accomplishment tends to be largely overlooked, even in research like that mentioned above does tend to take a basic position assuming the constructive ordering of social life with / around technologies. This core mission of EMCA feels double edged enough to potentially both act as a complement for this developing wealth of more critical 'social implications of AI' focussed research, yet at the same time offer a strong deflationary treatment of AI-in-the-world by focussing on overlooked mundanities of human-AI interaction that may help teach technologists once again—since agentification constantly works to make everyone forget—that social interaction is always a much harder environment than they imagine.

To conclude, although works by Suchman, Button et al. or Shanker necessarily form key guides in our filling out EMCA's promise and approach to AI-in-the-world, I'd like to think there is still a huge exploration of accomplishment still left uncharted.

**What's EMCA To Do? Evangelism, Service, Hybridity, Critique**
As a practical matter, what might EMCA want to do here? Our research network has discussed this. One such mode is 'evangelism'—of which my rendering is: attempting to enthuse AI-oriented researchers of the potential value of adopting EMCA approaches, or perhaps in them embracing particular concepts. Another is 'service'—which I take to mean working alongside technologists to provide insights derived from EMCA studies, rather than expecting conceptual or methodological impacts as per the evangelical mode. The difference mainly seems to be about a) creating change in the other, or b) moving alongside the other complementarily. There is a third version which is perhaps some mix of both service and evangelism (note that they aren't mutually exclusive). A fourth option might be EM hybridity—whereby EM and maybe CA researchers adopt technical skills themselves and through becoming competent with AI technologies from technologists' perspectives, then provide EMCA oriented accounts of those technical practices. There is a world in which EM hybridity can speak to STS, sociology, but also potentially if crafted appropriately, as studies for HRI, HCI, and other human- or social-oriented venues which have AI contingents or interests. Fifthly there is maybe an extension of Button et al.'s mode, that of conceptual critique—perhaps playing on emergent concepts in AI discourse like 'explainable AI' or 'responsible AI' and so on.

What does the evangelism-orientation hope to achieve? Possibly this could range from wider adoption of EMCA through to something more modest, like introduction of novel concepts or respecification of existing AI concepts. I feel that Moore and Arar's (2019) book *Conversational UX Design* tends to attempt this in quite an explicit way, building EMCA inspired work into a design framework. This work, in my view, attempts to contribute something that might be recognised as 'theory' by target disciplines (certainly this is the case for HCI). A service relationship orientation by comparison tends to foreground 'study-ness': that is, they are founded in producing

*relevant EMCA studies* as their primary task, and secondarily *translating* their outcomes to disciplinary interests. An example of this tendency might be found in studies of social robotics by Pitsch et al. (2016) or Pelikan et al. (2020)—to name some prominent exemplars. I detect the aim here is less to *teach* EMCA and rather focus more or less explicitly on providing 'design implications' or at least something like 'lessons'. (Work I've been involved in too perhaps fits more into the service mode.) At the same time this distinction does not feel entirely binary as studies like these often also have ambitions for conceptual (re)specification and some moderate advocacy of EMCA. Maybe it is all in the framing, where the strength of such advocacy sets up what kind of thing we are dealing with.

Now, some other caveats and concerns emerge at this point.

**Bitter Experience Has Taught Me Well**
There is a potential as an EMCA researcher for more technology-oriented research(ers) to eat you up and spit you out. Your research will be misunderstood, mischaracterised, misinterpreted, or maybe just ignored. This is certainly not unique to EMCA research, but it still bears repeating. EM and CA work could well receive a strong welcome, but ultimately few will probably appreciate your underlying philosophical motivations simply because—understandably—that is just not a priority for them. Instead EMCA may well be seen as just another fashion or trend in research methods for understanding human-AI interactions. Or perhaps EMCA will be seen as a service in aid of technology-driven research, the agenda of which in turn really will dictate 'what matters' and 'where we go'. Technological solutionism (Morozov 2014) is a strong force that still influences big chunks of fields like HCI in spite of many challenges mounted against this ideology. EMCA in service of solutionism is going to encounter trouble (see 'design' below). Somewhat similarly, methodological innovation has become a key staple of HCI-affiliate research (and I'd expect fields with some cultural connection to HCI like HRI to probably mirror this in some way)—so this means part of the cultural currency one gathers as a researcher *is* by 'moving on' which means *your research* is now passé. This sounds quite alarming, maybe, but I feel this is what has happened in HCI—and certainly not just to EM and CA oriented research only.

On this point, there is a slightly more optimistic take which I outline next.

The prospects for any affiliation between EMCA and AI (including AI systems design) could be compared with that between EMCA and HCI. It's worth recalling some recent past here, if only to avoid repeating history. For HCI, a connection with EM in particular goes back at least to the 1980s, most prominently in Suchman's (1987) very well-known critique of the cognitive modelling of users in expert systems ([to operate photocopiers—this is Xerox, remember](#)). Of key relevance to us as EMCA researchers prodding at 'AI', Suchman's work cut across multiple currents of then-contemporary AI and HCI, providing a powerful critique of GOFAI techniques and their assumptions, revealing how AI researchers (in this case Newell and others) had redefined HCI design problems into different, easier problems that bore little relation to how interaction actually unfolded. (Sidenote: I am repeatedly reminded of Suchman's work whenever EMCA research touches on AI, and often wonder if we are really providing something more substantive beyond what she already showed so clearly.)

One side effect of Suchman's work was the emergence of all sorts of disciplinary challenges that should concern us now for EMCA's deployment for understanding sociotechnical AI systems. Firstly: Suchman's work has often been misunderstood or mischaracterised. It has been continually confused with ethnography even though it makes no claims for this (Rooksby 2013), probably due to Suchman's position as an anthropologist by background. And it has been subject to theoretical capture by HCI researchers recasting it as "situated action theory" (check any HCI textbook), thereby rendering it a docile object that may sit calmly in the panoply of HCI theories (which are pretty much all 'borrowed'). And of course, Suchman's EM-driven points have also—perhaps inevitably—been subject of mischaracterisation by cognitive science (see Vera & Simon (1993), and Suchman's response (1993)).

This has all led to a characterisation of ethnomethodology in particular and to a lesser extent conversation analysis (due to EM being the focal interest of relevant actors in HCI) as being something like 'a kind of ethnography'. In a sense this is not a technique of dismissal but rather a disciplinary pigeonholing that one could find uncomfortable, unhelpful, or misrepresentative. Further, this impression has not been helped by in HCI research communities by some of the positions taken by a few EM proponents in HCI (see Crabtree et al. 2009). Disciplinary trouble has also been inflamed by an implied disdain or lack of appreciation for other approaches in HCI on the part of certain EM researchers in HCI, adopting a particular kind of 'evangelism' mode—not that 'evangelism' necessarily *must* be this way but it does offer a warning about the *style* adopted and how such an approach could slide into something like dismissal or even intellectual chauvinism. The net result is that the legacy of EM in HCI is—in my humble opinion—resolutely viewed through the lens of the researchers originally performing that evangelism, resulting in some significant confusions, conflations, and some very unusual renderings of EM itself (see Blackwell et al. 2017 for example), much of which may be attributed to (equally unusual) refusals to carefully distinguish between ethnography and ethnomethodology when broaching EM's historical role, and potential future role, in HCI. If you call what you are doing 'EM-informed ethnography' to help provide clarity of position, it does not then help to turn around and complain that other kinds of ethnography aren't like yours. Dourish has written pointedly on this matter (Dourish 2014, p14).

I could go on, but instead I'll briefly summarise some other potential sticking points which EMCA's approach to AI could learn from HCI:
- EMCA understood 'as a method' (therefore bringing it into commensurability with other methods; this has happened in HCI);
- Assumption of your centre also being others' centres of interest (design, technical development, theory, activism, etc.);
- Made the same as or affiliated with ethnography and ethnographic approaches including anthropology (i.e., shouldering others' baggage, when you have plenty of your own already);
- Collected together as 'interactionist', placed on a spectrum of disciplinary organisation between the 'macro' and the 'micro' (e.g., micro-ethnography);
- Having to deal with assumptions of compatibility with constructive analysis.

A possible aggravating factor for both the above could also have been the long documented tension between EM's work and the significance of HCI's design-orientation (Hughes et al. 1992). For me, it is this focus on *design*—"how things ought to be" in Simon's terms (1996)—that distinguishes HCI from other fields investigating humans and technology. Such tensions are certainly not exclusive to EMCA; see Dourish's (2006) paper on "implications for design" which spells out this problem clearly in terms of the role of ethnography in HCI. Why is this relevant? I believe a design-orientation tends to be present in HCI-adjacent fields like HRI, but also I feel it likely is widespread in AI-affiliates too. EMCA researchers might do well to remind themselves of this when trying to connect with human-AI oriented research. This question of design and its relationships will, I think, configure the modes of operation outlined above—certainly the 'service' relation most obviously. Design has always been a tricky prospect for EM in HCI for several reasons. Much conflict is driven simply by different purposes: design-orientation is about finding solutions to design problems or creating new design possibilities. It is a 'leap' to go from studies to design because the studies pursue their own particular logics and needs, configured by the necessities of disciplinary relations, interests of the EMCA researchers in question, etc. Whether such studies 'map' clearly to design outcomes (e.g., particular solutions, new possibilities, etc.) is not guaranteed. Filtering and self censorship of direction based in *optimising* outcomes for design could be a danger here as noted by Dourish. There is a question about what such studies 'do' for EMCA and whether they become subordinated to technologists. Technologists always have more money thrown at them, so there is a natural power imbalance. With such power relations emerging between researchers, their fields, etc. the introduction of (technical) design interests can easily begin dictating terms.

Finally, I refer back to Anderson and Sharrock's paper "Has ethnomethodology run its course?" (2017) where they provide further provoking questions of EM's potential purpose and future relationships with disciplines (and its own programme):

> "What are the methodological principles which facilitate or prohibit EM's triangulation on the idealisations which form the equivalent for other disciplines of its own re-specifications of the praxeological rule in sociology? Do or can these principles allow a straightforward exporting of EM findings and descriptions into those disciplines? Or is the parallel with sociology a general one with EM consistently constituting itself as asymmetric to and incommensurable with any discipline whose topics its takes up?" (p. 20-21)

**Moving Beyond EM and CA *for* Design**
It's possible a misstep of some EM and CA work in HCI was to act in a way too beholden to design. There are good reasons for this capture, and I don't wish to suggest the protagonists should have known better. There was much low-hanging fruit in the 80s, 90s and 00s. But hindsight might give us some better ideas about how *close* to get to design in less well-trodden fields for EMCA like HRI, for instance. Maybe there is room for the 'fourth way' I noted previously—something like hybrid studies—*which takes AI's relationship to design as its phenomena*? So, an ethnomethodological interest in those other approaches present in HCI as social phenomena worthy of investigation might also be productive for EMCA. Thus we could examine those methods and procedures applied to AI research in human-AI oriented fields. By way of example, we recently wrote an EM paper on Wizard of Oz methods for AI systems (Porcheron et al. 2020), which simultaneously attempted to reflect upon human-AI interaction (in this case voice activated, 'smart' robotic vacuum cleaners), but also to disassemble some of the *methods of HRI* upon which elements of its research programme is built. In doing so we can begin to understand the relationship between those methods and what is produced as relevant 'findings' for a given disciplinary endeavour.

There are also some more fundamental questions raised for any EMCA encounter with AI; whether the discourse, the communities, or the concepts. I wonder about EM's sense of unique adequacy and how AI technologies are to be handled. What is clear is that EM and CA—in my view—should be cognisant of and exercise vigilance about the 'constitutive' nature of AI, that is, many constitutive elements are erased in the production of AI accounts, such as the work to make the AI work (Mair et al. 2021). Should we only be oriented by participants' understandings? If we know more about underlying technologies does that impact the phenomena for us? To engage seriously with AI-affiliate fields how much technical understanding is necessary?

**Conclusion: A Reminder on Incommensurability**
Much of the inspiration for writing this was reading Wieder's (1993) paper on commensurability. In it he argues that the very basis for a meaningful exchange of ideas may be fundamentally hobbled by incommensurability between parties. If intersecting programmes are fundamentally at odds—for instance, there is rarely *any* questioning of the cognitive basis of AI as driving ideology—then they may well hit a stalemate. How can you 'go on' with people who have incommensurate conceptualisations of sociality and social organisation? Should EMCA ever get caught up in classic AI discussions? Like if computers can 'understand', have 'self awareness', 'consciousness' or 'reasoning'? What about more 'simple' things like 'reading' or 'hearing'? How far do you go—should you go—to bridge differences, to attempt to overcome this incommensurability? How much can / should we bracket off to make incommensurability tractable? You might find it easier to bracket off and unpack AI discourse about 'can machines think?' than the discourse of machines doing 'parsing'. The latter takes familiar concepts then turns them into technical ones. Then—like 'learning' or 'recognition'—these technical concepts get adopted back and blend with their borrowed everyday sense of the term, violating the incommensurate nature of the jumps. Finally, how should EMCA research position itself so as to navigate these different language games?

**Acknowledgements**
Much of this discussion was inspired by the EMCA research network on AI, organised primarily by Saul Albert, Kati Cyra, Hannah Pelikan, but featuring a cast of many others. I am grateful for the feedback on this piece from the network. This research is supported by the Engineering and Physical Sciences Research Council [grant number EP/T022493/1].